\documentclass[8pt]{article}
\usepackage{amssymb}
\usepackage{times}
\usepackage{graphicx}
\usepackage{sectsty}

\begin{document}

\sectionfont{\large}

\begin{center}
{\Large \bf Exclusive $J/\psi$ production in ultraperipheral Pb+Pb collisions \\
to NLO pQCD}
\vspace{0.5cm}

{\large K.~J.~Eskola, C.~A.~Flett, V.~Guzey, T.~L\"oyt\"ainen\footnote{Speaker; presented at the \textit{Quark Matter 2022} Conference, Krak\'ow, Poland, April, 2022}, H.~Paukkunen}
\vspace{0.5cm}

{University of Jyvaskyla, Department of Physics, Finland\\
Helsinki Institute of Physics, University of Helsinki, Finland}

\vspace{0.5cm}

\end{center}

\begin{abstract}
We present the first NLO pQCD study of coherent exclusive $J/\psi$ photoproduction in ultraperipheral heavy-ion collisions (UPCs) at the LHC. Taking the generalized parton distributions (GPDs) in their forward limit, as parton distribution functions (PDFs), we quantify the NLO contributions in the rapidity-differential cross section, show that the real part of the amplitude must not be neglected, study the gluon and quark contributions, chart the scale-choice and PDF uncertainties, and compare the NLO results with LHC and HERA data. We show that the scale dependence is significant but a scale choice can be found with which we reproduce the 2.76 and 5.02 TeV UPC data. In particular, we show that the process is clearly more sensitive to the nuclear quark PDFs than thought before.
\end{abstract}
  
\section{Introduction}
It was originally proposed by Ryskin in the context of leading-order pQCD and collinear PDFs \cite{Ryskin:1992ui} that coherent exclusive $J/\psi$ photoproduction off protons, $\gamma$ + $p$ $\rightarrow$ $J/\psi$ + $p$, is a promising probe of the gluon PDF of the proton: The forward cross section $d\sigma/dt(t=0) \propto (xg(x,Q^2))^2$ where $x= M_{J/\psi}^2/W^2$ and $Q^2=M_{J/\psi}^2/4$, with
$W$ the center-of-momentum-system (c.m.s.) energy and $M_{J/\psi}$  the $J/\psi$ mass.
Subsequently, coherent exclusive $J/\psi$ photoproduction in UPCs at the LHC, Pb + Pb $\rightarrow$ Pb + $J/\psi$ + Pb, has then  been suggested to efficiently probe the nuclear gluon distributions, see e.g.~Refs.~\cite{Adeluyi:2011rt,Guzey:2013qza}. For LHC data, see Refs.~\cite{ALICE:2021gpt,LHCb:2021bfl} and those in \cite{Eskola:2022vpi}.
Until now, coherent $J/\psi$ photoproduction in $A$ +$A$ UPCs has been studied only to LO pQCD, although NLO studies for the $\gamma + p$ case exist \cite{Ivanov:2004vd,Jones:2015nna,Flett:2020duk}. Thus there is an obvious need for an extension to NLO for the UPCs, the results of which we have recently presented  in Ref.~\cite{Eskola:2022vpi} and are reporting now here.

\section{Theoretical framework}

In UPCs of nuclei $A_1$ and $A_2$, the coherent $J/\psi$ photoproduction cross sections differential in the $J/\psi$ rapidity $y$ can be computed as
\begin{equation}
        \frac{d\sigma^{A_1A_2\rightarrow A_1VA_2} }{dy} = \left[ k \frac{dN_\gamma^{A_1}}{dk} \sigma^{\gamma A_2 \rightarrow VA_2} \right]_{k=k^+} \\
        + \left[ k \frac{dN_\gamma^{A_2}}{dk} \sigma^{A_1 \gamma\rightarrow A_1 V}  \right]_{k=k^-} 
\label{XS_plus_minus}
\end{equation}
where $A_{1,2}=$~Pb,  $V=J/\psi$, $k^\pm(y)$ are the energies of the photon emitted by the nucleus $A_{1,2}$, and $dN_\gamma^{A_{1,2}}/{dk}$ are the Weizs\"acker-Williams photon fluxes supplemented with a requirement of having no hadronic interactions. Here 
\begin{equation}
\sigma^{\gamma A_2 \rightarrow VA_2} = \frac{d\sigma_{A_2}^{\gamma N \rightarrow VN}}{dt} \bigg|_{t=0} \int\limits_{t_{\rm min}}^\infty dt' |F_{A_2}(-t')|^2, \label{eq:xsec1}
\end{equation}
where  $t$ is a Mandelstam variable, and the nuclear form factor $F_{A_2}$ is obtained as a Fourier transform of the Woods-Saxon nuclear density distribution (and correspondingly for $\sigma^{A_1 \gamma\rightarrow A_1 V}$). The $t$-differential cross section is given by the square of the collinearly factorized per-nucleon amplitude \cite{Collins:1996fb},
\begin{equation}
\mathcal{M}^{\gamma N \rightarrow VN}_A \propto \sqrt{{\langle O_1 \rangle_V }}
\int\limits_{-1}^1 dx [ T_g (x,\xi) F^g (x,\xi,t,\mu_F) 
    + T_q (x,\xi ) F^{q,S} (x,\xi,t,\mu_F) ] \nonumber
\end{equation}
where ${\langle O_1 \rangle_V }$ is the NRQCD element given by the ${\cal O}(\alpha_s^2)$  leptonic decay width of $J/\psi$ \cite{Ivanov:2004vd}, $T_{g,q}$ are the pQCD coefficient functions calculated to ${\cal O}(\alpha_s^2)$ in Ref.~\cite{Ivanov:2004vd}, $\xi$ is the skewedness parameter, and $\mu_F$ the factorization scale. The GPDs,  $F^{g}$ for gluons, and $F^{q,S}$ for the quark singlet, are here taken in their forward, no-skewedness, limit,
\begin{eqnarray}
F^{g} (x,0, 0,\mu_F) &=& F^{g} (-x, 0 , 0,\mu_F) = xg(x,\mu_F) \,, \nonumber \\
F^{q,S} (x,0,0,\mu_F) &=&  u(x,\mu_F) + d(x,\mu_F) + s(x,\mu_F) + c(x,\mu_F) \, \\
F^{q,S} (-x,0,0,\mu_F) &=& -\bar u(x,\mu_F) - \bar d(x,\mu_F) - \bar s(x,\mu_F) - \bar c(x,\mu_F)\,, \nonumber 
\end{eqnarray}
where $x>0$. Note that quarks contribute only at NLO here. For the nuclear PDFs (nPDFs), we use EPPS16 \cite{Eskola:2016oht}, nCTEQ15 \cite{Kovarik:2015cma} and nNNPDF2.0 \cite{AbdulKhalek:2020yuc}. Throughout this work, we set the renormalization scale $\mu_R$ to be equal to $\mu_F$. We have also cross-checked our numerical results with two different methods for doing the complex integrations. For more details, see \cite{Eskola:2022vpi,Ivanov:2004vd}.

\section{Results}
The main results from our NLO study are collected in Figs.~1-5. 
Figure 1 shows that the scale dependence is considerable but an ``optimal" scale $\mu=0.76 M_{J/\psi}$ can be found with which the LHC data are well reproduced. 
\begin{figure}[!]
\centering{
\includegraphics[width=0.73\textwidth]{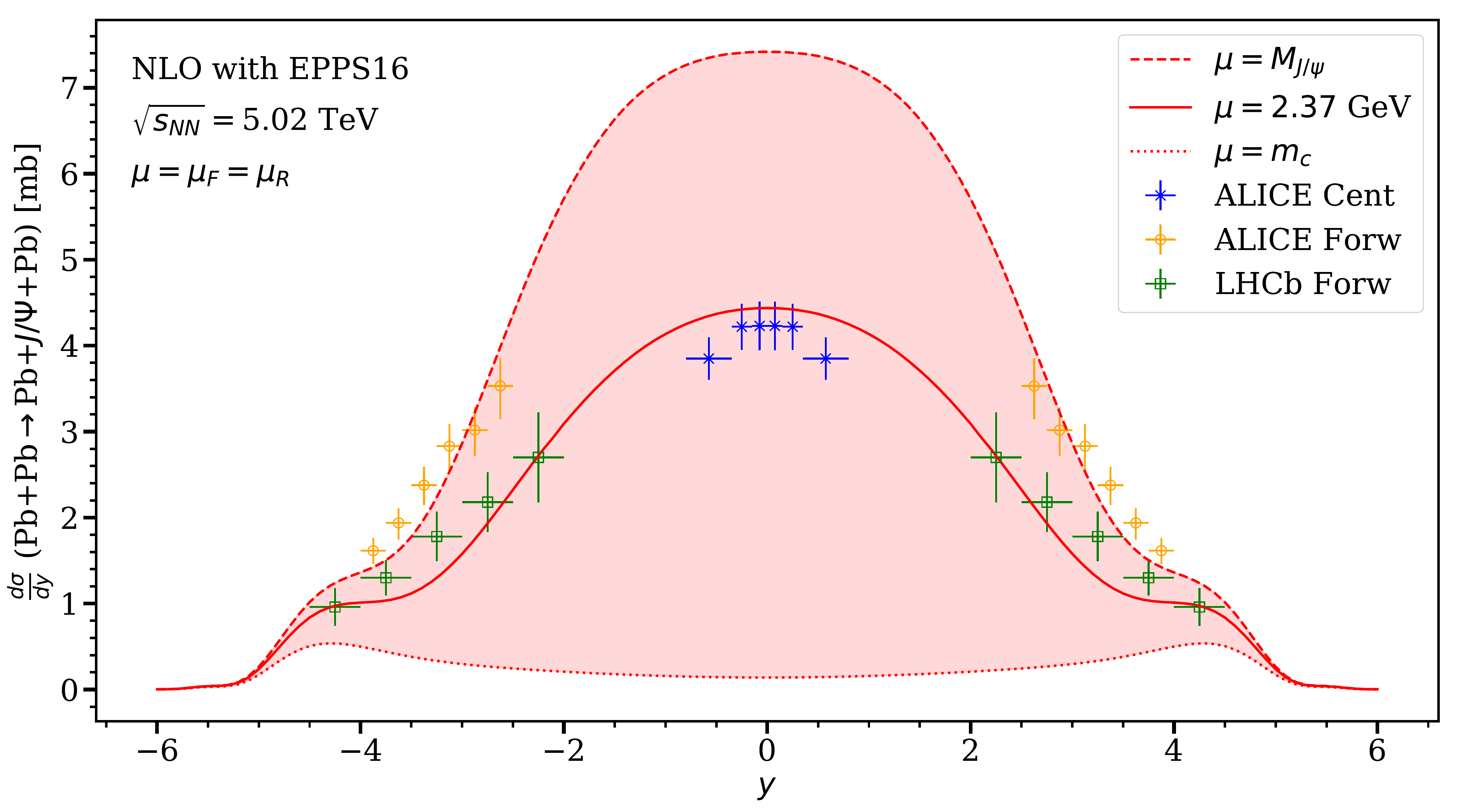} \\

\vspace{-0.1cm}\hspace{-0.3cm}
        \includegraphics[width=0.75\textwidth]{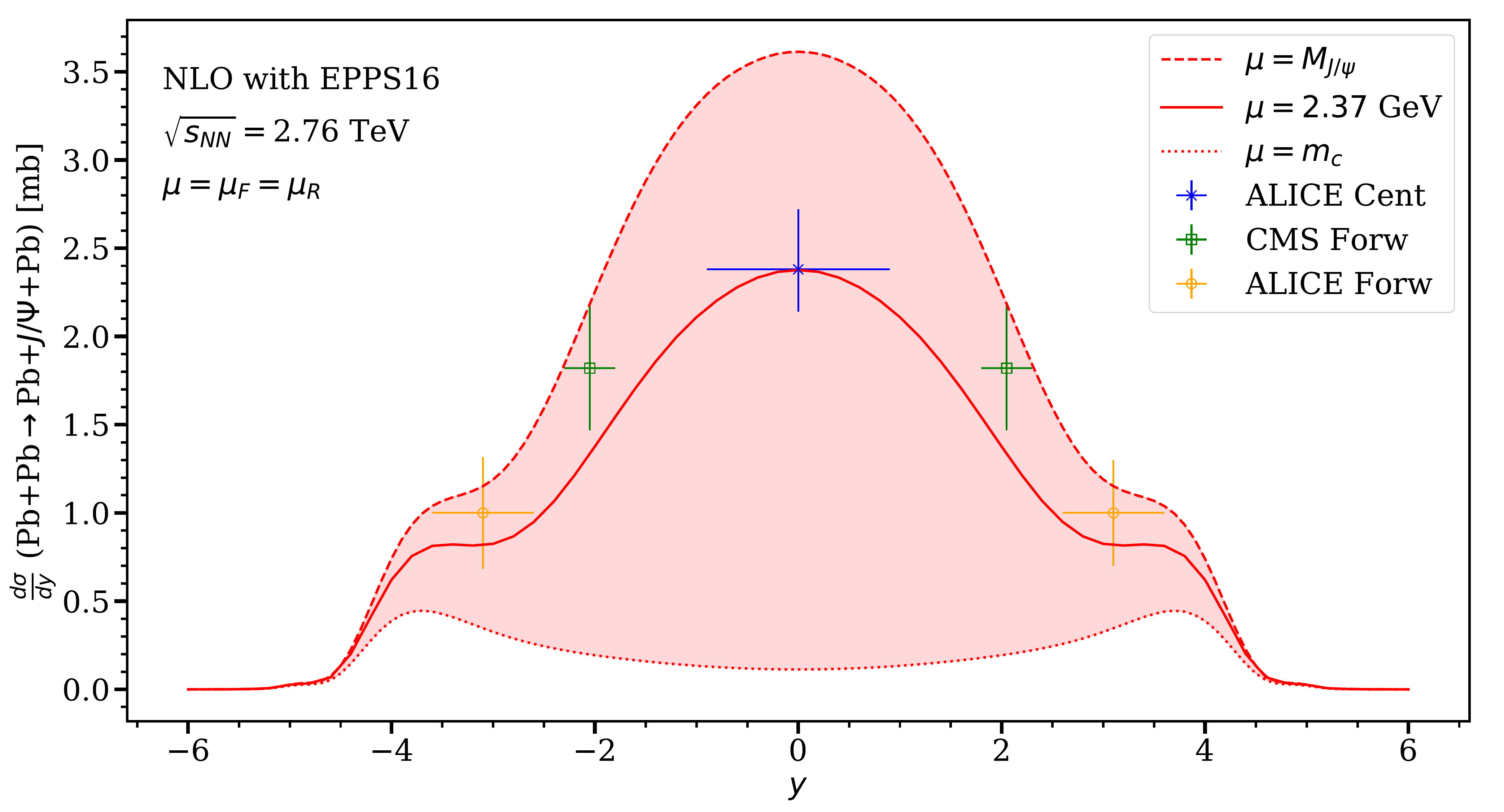}}
\vspace{-0.3cm}
				\caption{\small Rapidity-differential coherent exclusive $J/\psi$ photoproduction cross section vs.~rapidity in Pb+Pb UPCs at $\sqrt{s_{NN}}= 5.02$ TeV (upper panel) and 2.76 TeV (lower panel), computed with EPPS16 nPDFs, and scales $\mu = M_{J/\psi}/2,\,0.76 M_{J/\psi}$ and $M_{J/\psi}$. For the references to the LHC data shown, see \cite{Eskola:2022vpi}. Figure from \cite{Eskola:2022vpi}.}
\vspace{-0.2cm}
\end{figure}
The same ``optimal" scale works reasonably well also for the $\gamma+p$ baseline, as seen in Fig.~2.   
\begin{figure}[!]
\centering{
\includegraphics[width=.76\textwidth]{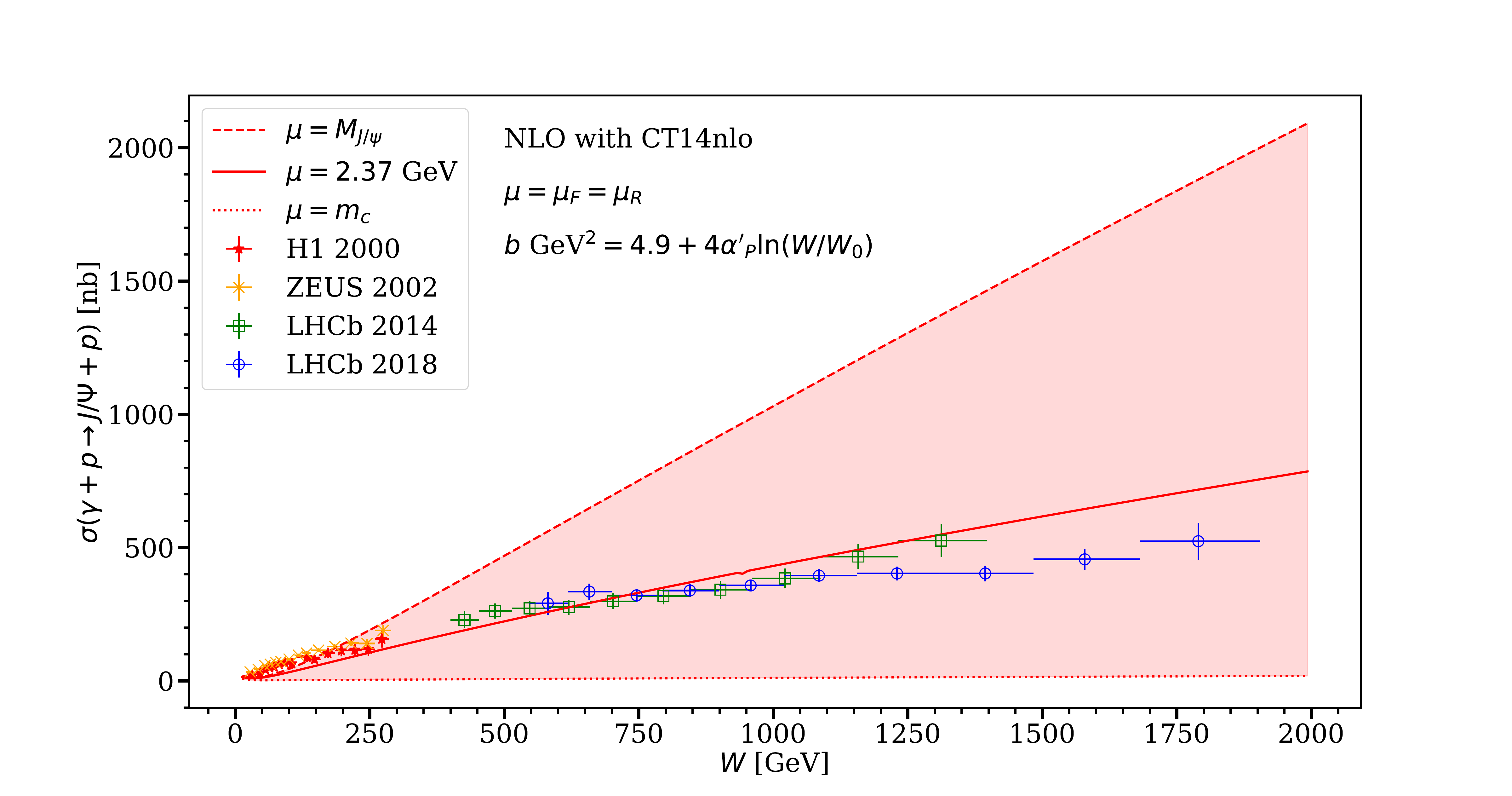}	\hspace{0.5cm}			
}
\vspace{-0.3cm}
\caption{\small Exclusive $J/\psi$ photoproduction cross section in $\gamma+p$ collisions vs.~c.m.s.~energy $W$, computed with CT14NLO PDFs \cite{Dulat:2015mca} and scales $\mu/M_{J/\psi} = 2$, 0.76 and 1. For the references to the HERA and LHC data shown, see \cite{Eskola:2022vpi}. From \cite{Eskola:2022vpi}.}
\vspace{-0.5cm}
\end{figure}
Figure 3 demonstrates the complex structure of the NLO cross section which in Pb+Pb UPCs results from an interplay between the pQCD cross section, photon fluxes from both nuclei, and the nuclear form factor. The upper panel shows that unlike in LO where the imaginary part of the amplitude clearly dominates, in NLO the situation becomes more involved and the real part cannot be neglected. The lower panel shows that in NLO, the \textit{quark} contribution dominates at $y=0$ -- perhaps the most striking result of this study. This follows from the canceling LO and NLO gluon amplitudes, as analysed in detail in Ref.~\cite{Eskola:2022vpi}. The ``shoulders" in the full NLO result arise because the NLO terms weaken the $W$ dependence of the pQCD cross section at small values of $W$.
Figure 4 shows the NLO cross sections computed with different nPDFs. While  
nCTEQ15 gives very similar results as EPPS16, nNNPDF2.0 deviates from them considerably. This is due to the very rapidly growing  small-$x$ gluon distributions in nNNPDF2.0. Finally, the propagation of the PDF uncertainties is shown in Fig.~5. The EPPS16 and nCTEQ15 nPDF uncertainties remain moderate, while the CT14NLO error bands are large and dominated by one error set where again the small-$x$ gluons rise very rapidly. 

\begin{figure}[!]
\centering{
 \includegraphics[width=.73\textwidth]{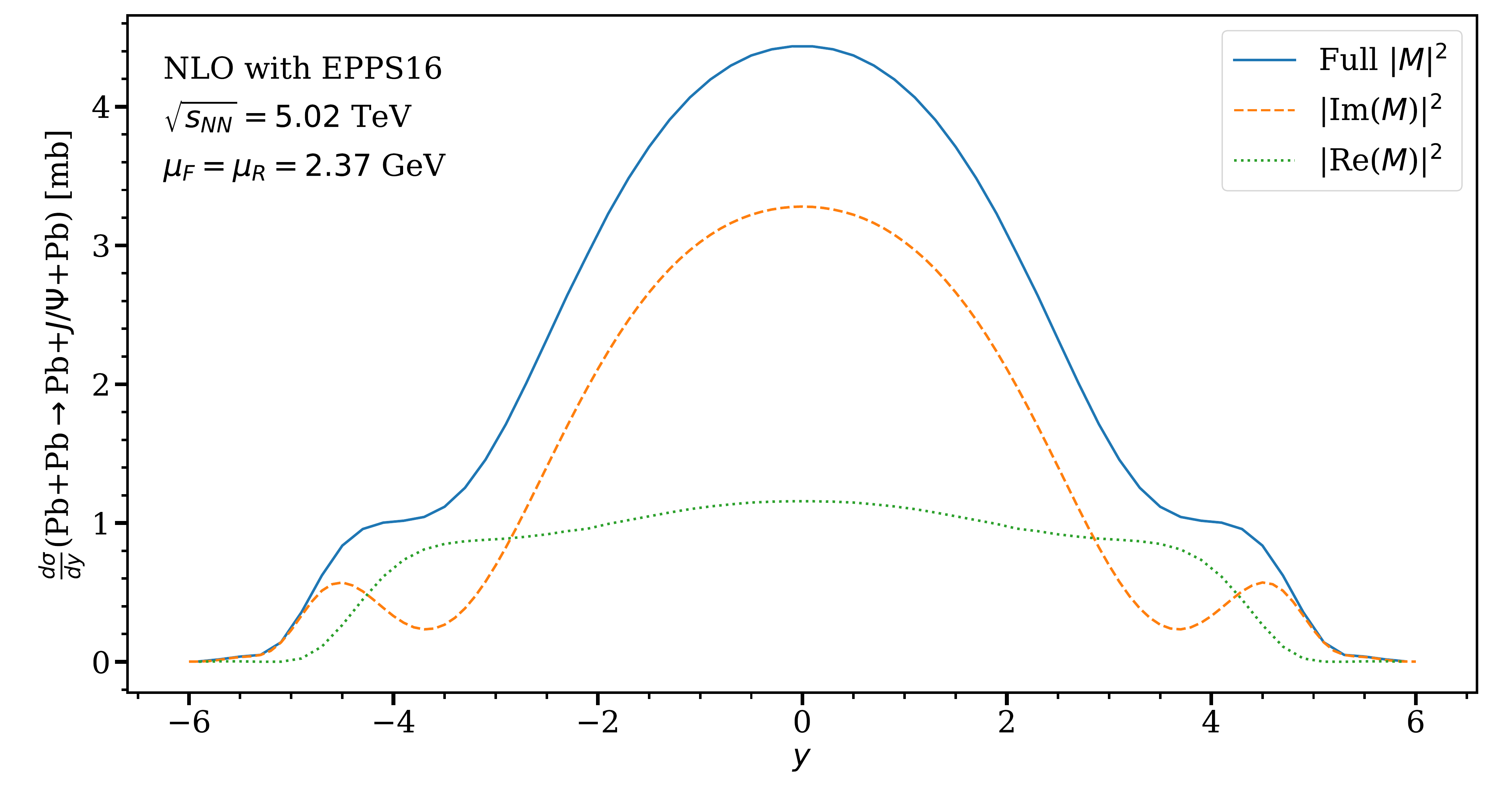}

\includegraphics[width=.74\textwidth]{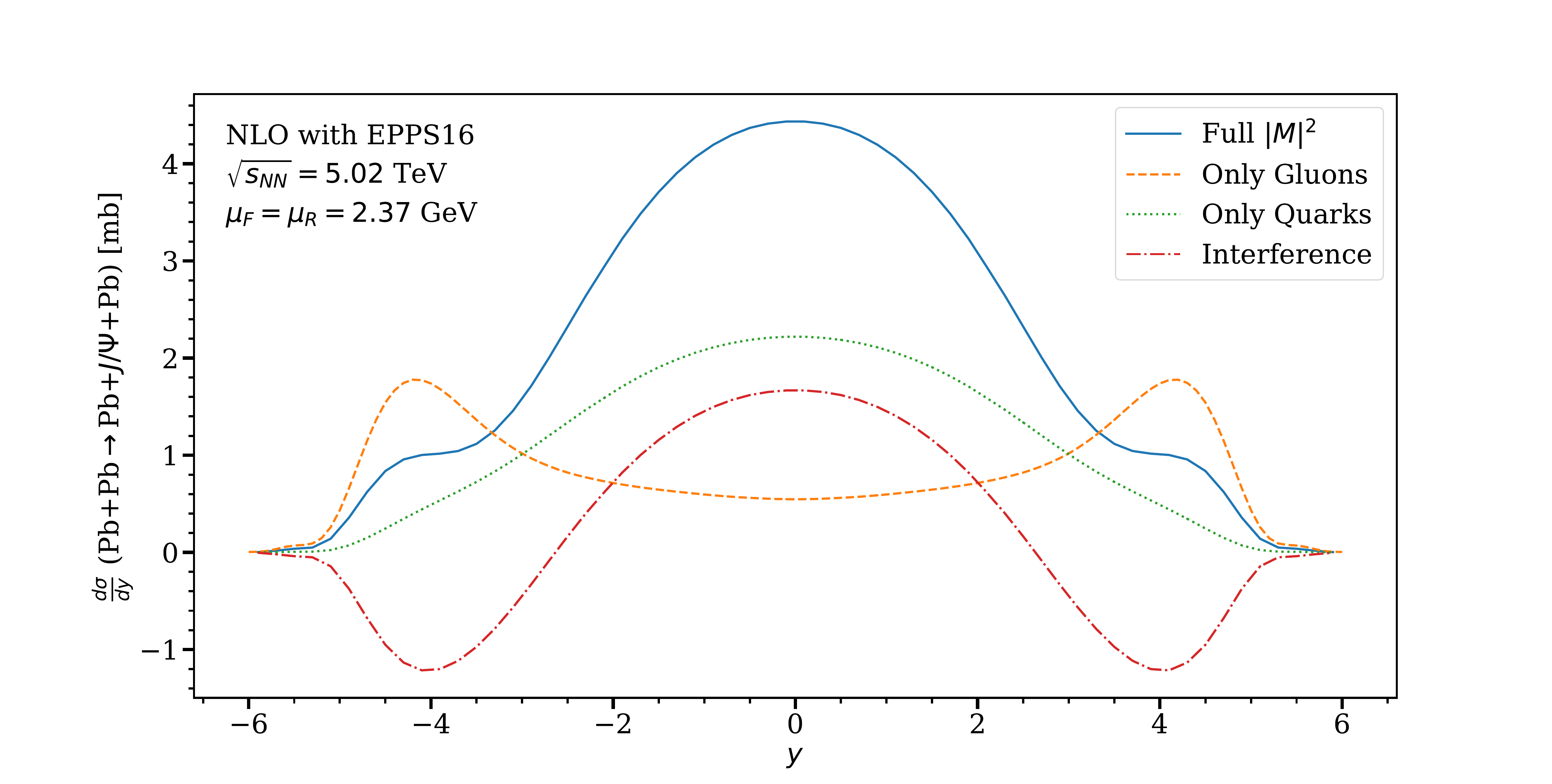}
}
\vspace{-0.3cm}
\caption{\small \textit{Upper panel:} Breakdown of the NLO cross section in the upper panel of Fig.~1 into contributions from the imaginary and real parts of the amplitude. 
\textit{Lower panel:} Contributions without quarks, without gluons, and from the quark-gluon interference terms alone.  Figures from \cite{Eskola:2022vpi}.
}
\vspace{1cm}
\centering{
\includegraphics[width=.73\textwidth]{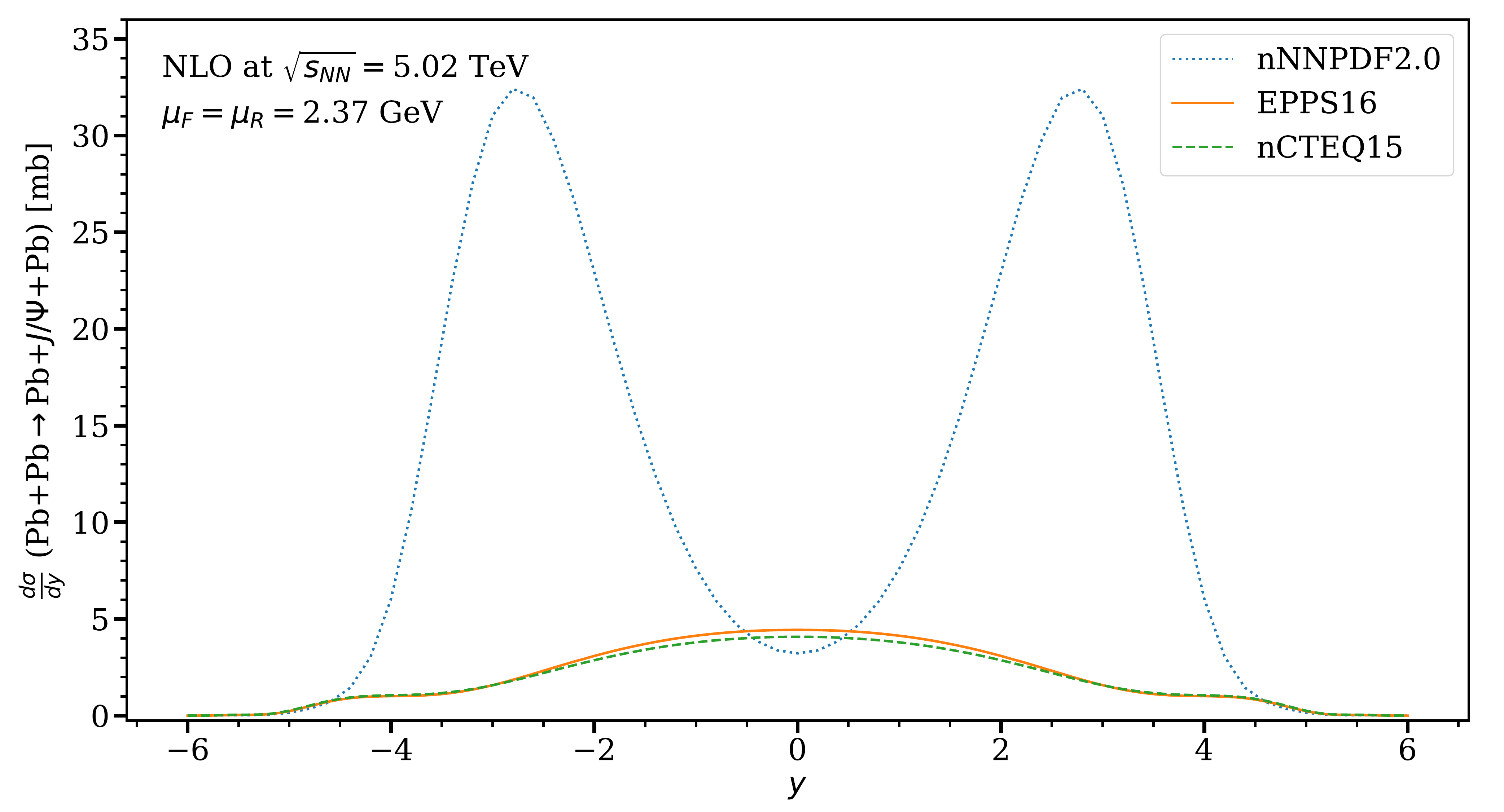}
}
\vspace{-0.3cm}
\caption{\small As Fig.~1 upper panel, but computed with three different nPDFs using the same ``optimal" scale. Figure from \cite{Eskola:2022vpi}.}
\vspace{-0.5cm}
\end{figure}

\begin{figure}[h!]
\centering{
\includegraphics[width=.79\textwidth]{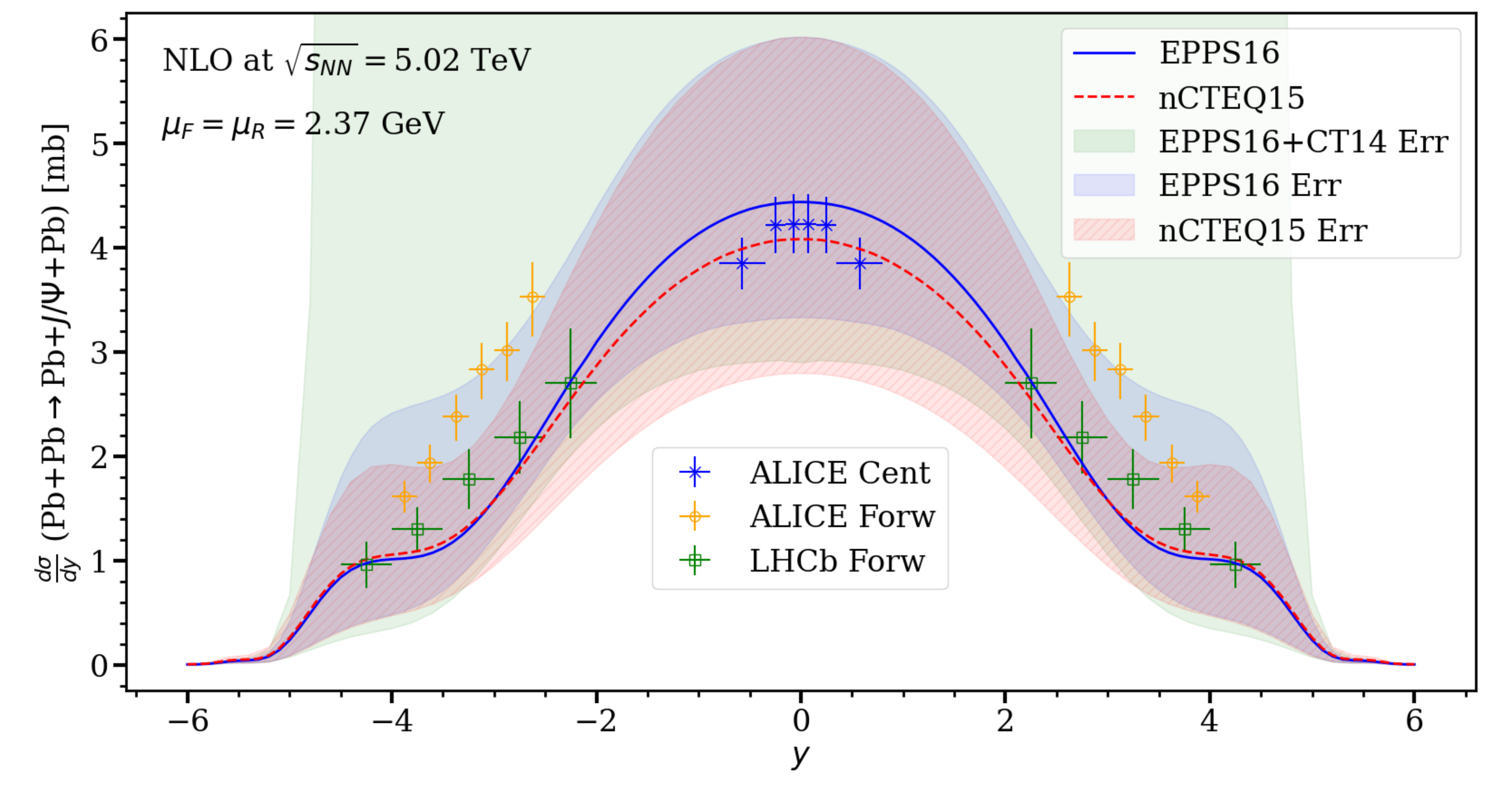}}
\centering{
    \includegraphics[width=.81\textwidth]{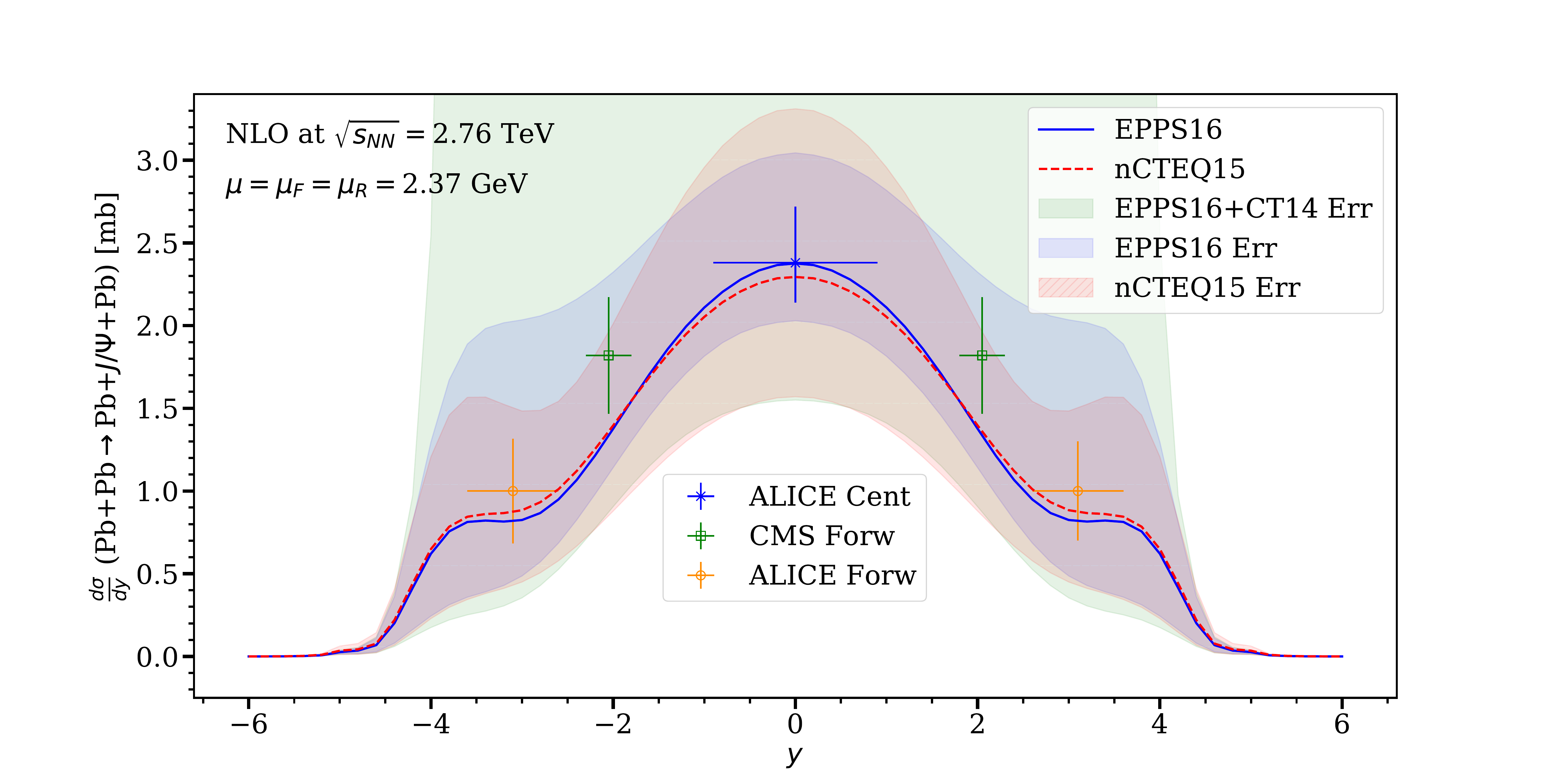}
}
\vspace{-0.3cm}
\caption{\small As Fig.~1 but with PDF uncertainties at the ``optimal" scale. From \cite{Eskola:2022vpi}.}
\end{figure}

\section{Summary}
We have presented the first implementation of the NLO pQCD cross section of exclusive $J/\psi$ photoproduction in Pb+Pb UPCs \cite{Eskola:2022vpi}.
In spite of the large scale-dependence envelope, there seems to be an ``optimal" scale with which we can, with EPPS16 and nCTEQ15, reproduce the LHC data well. We have shown that at NLO the real part of the amplitude must not be neglected, and that the canceling LO and NLO gluon amplitudes make quarks contribute much more than what was traditionally expected. Thus, the picture changes rather dramatically at NLO. We have also charted how the PDF uncertainties propagate into the computed cross sections.
In this exploratory study we assumed the forward limit for the GPDs.
In the future, it will be interesting to study the effects of GPD modeling. 

\vspace{0.5cm}
\noindent\textbf{\large Acknowledgements} 

\noindent We acknowledge the financial support from the Magnus Ehrnrooth foundation (T.L.), the Academy of Finland projects 308301 (H.P.) and 330448 (K.J.E.). This research was funded as a part of the Center of Excellence in Quark Matter of the Academy of Finland (projects 346325 and 346326). This research is part of the European Research Council project ERC-2018-ADG-835105 YoctoLHC.


\vfill

\begin{thebibliography}{99}
\vspace{-0.3cm}
\bibitem{Ryskin:1992ui}
M.~G.~Ryskin,
Z. Phys. C \textbf{57} (1993), 89
\vspace{-0.2cm}

\bibitem{Adeluyi:2011rt}
A.~Adeluyi and C.~Bertulani,
Phys. Rev. C \textbf{84} (2011), 024916
\vspace{-0.2cm}

\bibitem{Guzey:2013qza}
V.~Guzey and M.~Zhalov,
JHEP \textbf{10} (2013), 207
\vspace{-0.2cm}

\bibitem{ALICE:2021gpt}
S.~Acharya \textit{et al.} [ALICE],
Eur. Phys. J. C \textbf{81} (2021) no.8, 712
\vspace{-0.2cm}

\bibitem{LHCb:2021bfl}
R.~Aaij \textit{et al.} [LHCb],
JHEP \textbf{07} (2022), 117 
\vspace{-0.2cm}

\bibitem{Eskola:2022vpi}
K.~J.~Eskola, C.~A.~Flett, V.~Guzey, T.~L\"oyt\"ainen and H.~Paukkunen,
arXiv:2203.11613 [hep-ph]
\vspace{-0.2cm}

\bibitem{Ivanov:2004vd}
D.~Y.~Ivanov, A.~Schafer, L.~Szymanowski and G.~Krasnikov,
Eur. Phys. J. C \textbf{34} (2004) no.3, 297
[erratum: Eur. Phys. J. C \textbf{75} (2015) no.2, 75]
\vspace{-0.2cm}

\bibitem{Jones:2015nna}
S.~P.~Jones, \textit{et al.}, 
J. Phys. G \textbf{43} (2016) no.3, 035002
\vspace{-0.2cm}

\bibitem{Flett:2020duk}
C.~A.~Flett, \textit{et. al.}, 
Phys. Rev. D \textbf{102} (2020), 114021
\vspace{-0.2cm}

\bibitem{Collins:1996fb}
J.~C.~Collins, L.~Frankfurt and M.~Strikman,
Phys. Rev. D \textbf{56} (1997), 2982
\vspace{-0.2cm}

\bibitem{Eskola:2016oht}
K.~J.~Eskola, \textit{et al.},
Eur. Phys. J. C \textbf{77} (2017) no.3, 163
\vspace{-0.2cm}

\bibitem{Kovarik:2015cma}
K.~Kovarik, \textit{et al.},
Phys. Rev. D \textbf{93} (2016) no.8, 085037
\vspace{-0.2cm}

\bibitem{AbdulKhalek:2020yuc}
R.~Abdul Khalek, \textit{et al.},
JHEP \textbf{09} (2020), 183
\vspace{-0.2cm}


\bibitem{Dulat:2015mca}
S.~Dulat, \textit{et al.},
Phys. Rev. D \textbf{93} (2016) no.3, 033006
\vspace{-0.2cm}

\end{thebibliography}
\end{document}